\def\la{\;
\raise0.3ex\hbox{$<$\kern-0.75em\raise-1.1ex\hbox{$\sim$}}\; }
\def\ga{\;
\raise0.3ex\hbox{$>$\kern-0.75em\raise-1.1ex\hbox{$\sim$}}\; }
\documentstyle[psfig]{mn}
\begin{document}
\title[Monthly Notices:  D/H from absorption spectra]
{New aspects of absorption line formation in intervening turbulent clouds -- III. 
The inverse problem in the study of H+D profiles}
\author[S. A. Levshakov  et al.]
{Sergei A. Levshakov$^{1}$,
Wilhelm H. Kegel$^{2}$, and Fumio Takahara$^{3}$\\
$^{1}$Department of Theoretical Astrophysics, A. F. Ioffe Physico-Technical
Institute, 194021 St.Petersburg, Russia \\
$^{2}$Institut f\"ur Theoretische Physik der Universit\"at Frankfurt am Main,
Postfach 11 19 32, 60054 Frankfurt/Main 11, Germany\\
$^{3}$Department of Physics, Tokyo Metropolitan University, Hachioji,
Tokyo 192-03, Japan }
\maketitle
\begin{abstract}
A new method, based on a Reverse Monte Carlo technique and
aimed at the inverse problem in the analysis of interstellar (intergalactic) 
absorption lines is presented. 
We consider the process of line formation in media with a stochastic velocity field
accounting for the effects of a finite correlation length 
({\it mesoturbulence}).
This approach generalizes the standard {\it microturbulent} approximation,  
which is commonly used to model the formation of absorption spectra in
turbulent media.

The method allows to estimate from an observed spectrum 
both, the physical parameters of the absorbing gas, and 
an appropriate structure of the velocity distribution parallel to the line of sight.
The procedure is applied to 
a template H+D Ly$\alpha$ profile which reproduces the
Burles \& Tytler [B\&T] spectrum of the quasar Q 1009+2956 
with the DI Ly$\alpha$ line seen at the redshift $z_a = 2.504$.
 
The results obtained favor a {\it low} D/H ratio in this particular absorption
system, although the inferred upper limit for the hydrogen isotopic ratio
of about $4.6\times10^{-5}$ is slightly higher than that of B\&T 
[log (D/H) = $-4.60 \pm 0.08 \pm 0.06$, i.e. D/H$ \la 3.47\times 10^{-5}$].
This revision of the B\&T data in the framework of the mesoturbulent model
leads to limits on D/H consistent with standard big bang nucleosynthesis
[SBBN] predictions and observational constraints on both extra-galactic $^4$He mass
fraction Y$_p$ (Izotov {\it et al.} 1997) and $^7$Li abundance in the atmospheres of
population II (halo) stars (Bonifacio \& Molaro 1997).
SBBN restricts the common interval of the baryon to photon ratio $\eta$ for
which there is concordance between the various abundances to
$3.9\times10^{-10} \la \eta \la 5.3\times10^{-10}$. It implies that 
$0.014 \la \Omega_b h^2 \la 0.020$, 
where $h$ is the Hubble constant in units 100 km s$^{-1}$ Mpc$^{-1}$. 

\end{abstract}

\section{Introduction}

In two previous  papers (Levshakov \& Kegel 1997, and Levshakov, Kegel \& Mazets 1997, 
Paper~I and II hereinafter, respectively),
we investigated the formation of absorption lines in chaotic media
with finite correlation length of the velocity field
({\it mesoturbulence}), when the observed absorption spectra
correspond to only one line of sight. We have shown that
accounting for a finite correlation length in the large scale
stochastic velocity field changes the interpretation of absorption
spectra considerably. 
The line profiles, as well as the equivalent widths, may be strongly
affected by the correlations. 
Therefore the standard analysis of
absorption spectra based on the
{\it microturbulent} approximation and the 
Voigt profile fitting procedure may yield misleading results
concerning the physical parameters of the absorbing gas
(for numerous examples, see Levshakov \& Kegel 1994, 1996; 
Levshakov \& Takahara 1996a,b; Paper~I and Paper~II). 
At least one can say that the results obtained by the standard 
analysis are model dependent and, thus, are not unique.

In our cloud model we account for a stochastic velocity field
with finite correlation length $l$ but assume the absorbing
material to be homogeneous in density and temperature.
The model is fully defined by specifying 
the column density $N$, the kinetic temperature
$T_{kin}$, the ratio of the r.m.s. turbulent velocity to the thermal velocity
$\sigma_t/v_{th}$, the ratio of the cloud thickness to the correlation length
$L/l$, as well as one realization of the velocity field distribution
along the line of sight.
Because of the very large time scale for changes in the hydrodynamic flows
the random structure of the velocity field along a given line of sight
has to be considered as `frozen' over the exposure time. 
It follows that the observed absorption spectrum 
in the light of a point-like  source (star, QSO) 
corresponds to only one realization
of the velocity field. Therefore, the intervening absorption spectrum
cannot be considered as a time or space average in the statistical sense. 
This means that the actual distribution of the hydrodynamic
velocities at a given instant of time corresponds 
to an {\it incomplete} sample and, thus, may deviate from the average distribution
function assumed to be Gaussian.
Accordingly, significant deviations from the expected average
intensity $\langle I_\lambda \rangle$ can occur. -- For this reason we used in Paper~II
a Monte Carlo technique to calculate spectra corresponding to the absorption
along individual lines of sight.

In Paper~II we considered the direct problem, i.e. we specified
the physical parameters and generated individual
random realizations of the velocity distribution
[more exactly -- the distribution of the velocity component parallel to
the line of sight $v(s)$]
with which we then calculated individual spectra. The aim of the present
paper is the inverse problem, i.e. the problem to deduce physical parameters
from an observed spectrum.
To reproduce a given (observed) spectrum
within the framework of our model, one has to find the proper physical
parameters {\it and} an appropriate realization of the velocity field.
In general, $v(s)$ is a continuous random function of the coordinate $s$, 
but in the numerical procedure it is sampled at evenly spaced intervals $\Delta s$.
The necessary number of intervals depends on the values of
$\sigma_t/v_{th}$ and $L/l$, being typically $\sim 100$
for hydrogen absorption lines.
Thus, to estimate model parameters from the observed
spectrum one has to solve an optimization problem in a
parameter space of very large and variable (depending on $\sigma_t/v_{th}$, $L/l$) dimension.
It is known that such kind of problems may be solved using stochastic optimization methods.
The Reverse Monte Carlo [RMC] technique
based on the computational scheme invented by
Metropolis {\it et al.} (1953) proved to be adequate in our case. The method is
successfully used in many applications
where the state space of a physical system is huge
(see e.g. Press {\it et al.} 1992, or Hoffmann 1995).

We apply the RMC method to the problem
of determining the primordial deuterium abundance at high redshift.
In particular, the H+D Ly$\alpha$ profile observed by Burles \& Tytler
(1996, B\&T hereinafter) in the spectrum of the quasar Q 1009+2956 is considered.
This spectrum was selected since ($a$) it shows a pronounced
DI absorptions at the redshift $z_a = 2.504$, 
($b$) it was obtained with high signal-to-noise ratio and spectral resolution,
and ($c$) the total hydrogen column density estimated by B\&T from the
intensity level beyond the Lyman limit in this system is consistent with
the normalized intensities in the HI Ly$\alpha$ wings.

In the present paper we extend our study of the accuracy of the
D/H ratio determinations from our Galaxy (Paper~II)
to the very distant Ly$\alpha$-systems (putative intervening galaxies). 
The background source is supposed
to be point-like and even in the case of distant QSOs
we consider absorption along one line of sight only.
This means that any gravitational focusing which may, in principle, lead to an additional spatial
averaging over  different lines of sight
(an example may be found in Frye {\it et al.} 1997) 
is not taken into account. (The average
mesoturbulent H+D Ly$\alpha$ spectra have been considered by Levshakov \& Takahara 1996a.) 

In Section 2 a 
description of the RMC method and the procedure to test its validity
are given, while in Section 3  the method is applied to analyze 
an observed H+D Ly$\alpha$ spectrum.
The results obtained are summarized in Section 4.


\section{The Reverse Monte-Carlo method}

The RMC method belongs to the class of stochastic optimization
algorithms developed to solve
optimization problems with a very large number of free parameters. 
Contrary to the standard Monte-Carlo procedure in which
random configurations of a given physical system are generated
to estimate its average characteristics, the RMC
takes an experimentally determined set of data 
and searches for a parameter configuration which reproduces the observational data.

The inverse problem is always an optimization problem in which an objective function
is minimized.
The objective function defined over a space of very large dimension 
is known to have many local minima. Therefore optimization methods based
on the choice of `minimization direction' (so-called gradient methods) 
may fail because of the possibility
of being trapped in one of such minima. 
The RMC approach allows to overcome this problem. It can
get over the local minima because it does not use any concept of `direction'.
Applied to the analysis of absorption spectra, the RMC method may be formulated
as follows. 

\begin{figure*}
\vspace{0.0cm}
\centerline{\vbox{
\psfig{figure=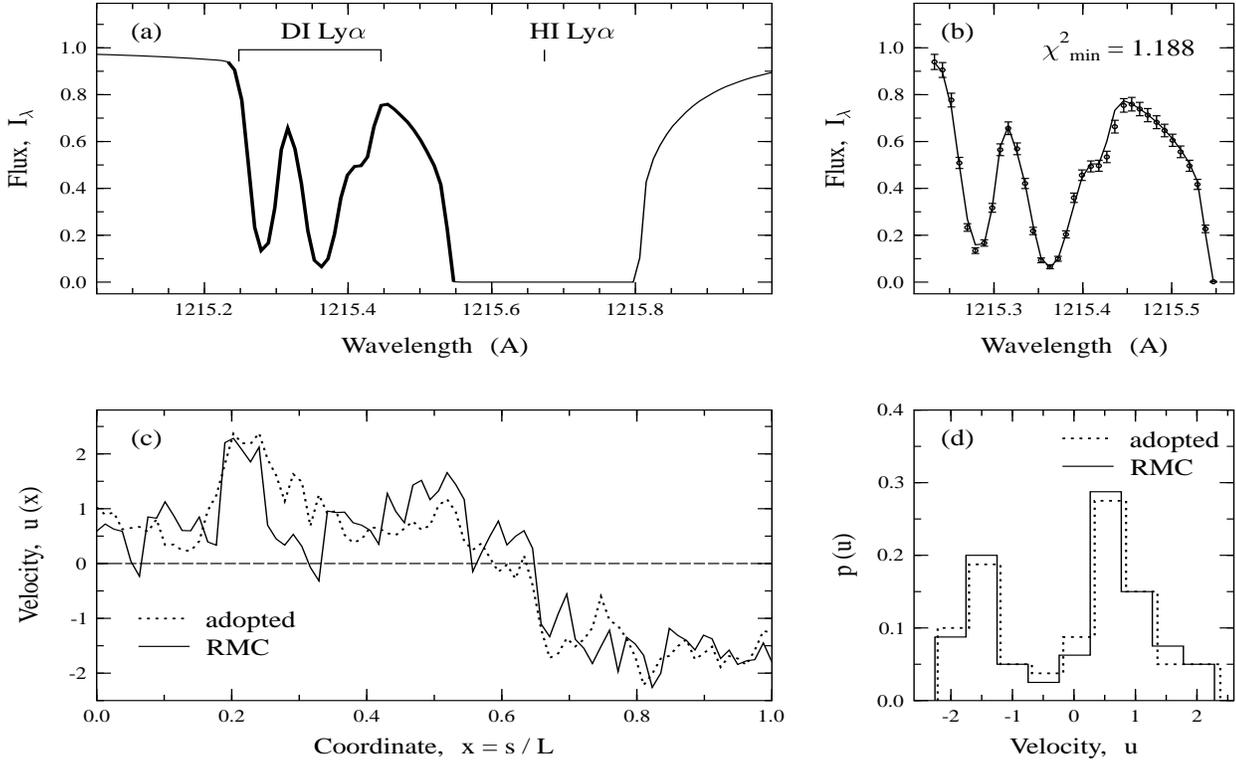,height=15.0cm,width=15.0cm}
}}
\vspace{-4.0cm}
\caption[]{
({\bf a}) -- A Monte-Carlo simulation of the H+D Ly$\alpha$
profile. The thicker curve marks the blue wing of the HI line and the
complex structure of the D absorption. ({\bf b}) -- The portion of the H+D profile 
marked in panel {\bf a}
is shown by open circles having error bars
corresponding to S/N = 30 at the continuum level, while the solid curve is 
the result of the RMC minimization.
({\bf c}) -- The adopted one-dimensional velocity distribution (dotted curve) and the
one determined by the RMC procedure (solid curve). ({\bf d}) -- Histograms $p(u)$ for the
adopted (dotted lines) and estimated (solid lines) projected velocity
distributions ($u \equiv v/\sigma_t$).
}
\end{figure*}

\subsection{Computational procedure}

The total parameter space consisting of the physical parameters (like
$N, T_{kin},$ etc.) and the velocity components $v_j$ parallel to the
line of sight at the individual positions $s_j$, is divided into two
subspaces.

Let $\Theta = \{\hat \theta\}$ represent the subspace of physical parameters
\{$N, T_{kin}, \sigma_t/v_{th}, L/l$\}. 
For the special case of H+D Ly$\alpha$\,  that we want to investigate, 
${\hat \theta}$ has to have one additional
component D/H -- the ratio of the DI to HI column densities.

Let further
$\{{\hat v}\} = \{v_1, v_2, \ldots, v_{k}\}$ be the vector of the velocity components
parallel to the line of sight at the spatial points $s_j$,
determined by the recurrent sequence (see Paper II)
\begin{equation}
v_j \equiv v(s_{j-1} + \Delta s) =
\xi \sigma_t \sqrt{1 - f^2} + f v_{j-1}\, ,\,   j = 2,3, \ldots ,k\, .
\label{eq:E1}
\end{equation}
Here 
\begin{equation}
f \equiv f(\Delta s) = \exp ( - |\Delta s|/l ) 
\label{eq:E2}
\end{equation}
is the correlation function
and $\xi$ is a random number picked from a normal distribution 
with $\langle \xi \rangle = 0$
and Var($\xi$) = 1.

For a chosen ${\hat \theta}$\, and velocity field distribution 
the objective function ${\cal L}$ is calculated according to
\begin{equation}
{\cal L} \equiv \chi^2 = \frac{1}{\nu} \sum^{m}_{i=1}
\left[ \frac{I(\lambda_i) - r(\lambda_i)}{\sigma_i} \right]^2\, ,
\label{eq:E3}
\end{equation}
where $r(\lambda_i)$ is the simulated random intensity [eq.(19) in Paper II],\, 
$I(\lambda_i)$ the observed normalized 
intensity within the $i$th pixel of the line
profile, $\sigma_i$ an estimate of the experimental error, and 
$\nu = m - n$ \,  the degree of freedom ($m$ is the number of data points
and $n$ is the number of fitted physical parameters, $n = 5$ in our case).

The minimization of $\chi^2$ is the main goal of the procedure.
Any parameter configuration with $\chi^2 \sim 1$ has to be considered as
a physically reasonable model for the interpretation of the observational
data.

The computational RMC procedure is split into two steps. At first, random values
for the physical parameters are chosen, i.e. the vector ${\hat \theta}$. Secondly,
an optimal velocity field configuration is estimated for these parameters.
The current $\chi^2$ value is computed and if it is larger than 
$\chi^2_{\nu,\alpha}$ (which
is the value of $\chi^2_{\nu}$ for a given credible probability
$P_{\alpha} = 1 - \alpha$) the whole procedure is repeated. In principle,
to choose the vector ${\hat \theta}$ in the physical parameter subspace
we could use any of the common optimization methods (like e.g. nonlinear simplex)
but we apply the same stochastic RMC procedure for both stages of the 
computational scheme. In detail it is described as follows.

\smallskip\noindent
{\bf 1}) A simulation box in the parameter subspace is specified
by fixing the parameter boundaries. 

\smallskip\noindent
{\bf 2}) ${\hat \theta}$ is chosen arbitrarily in the simulation box.

\smallskip\noindent
Next, the points {\bf 3}--{\bf 5} describe the evaluation of optimal velocity field
configurations for a given set of physical parameters.

\smallskip\noindent
{\bf 3}) A random realization of the 
velocity field is generated, and the corresponding value of 
$\chi^2$ is calculated.

\smallskip\noindent
{\bf 4}) Subsequently an element $v_j$ is chosen at random, and a random
change is given to this element in accord with  (\ref{eq:E1}).
If $j$ happens to be equal to 1, we put $v_1 = \xi \sigma_t$.

\smallskip\noindent
{\bf 5}) The new value of ${\cal L}$ is calculated.
If $\Delta {\cal L} < 0$, i.e. if $\chi^2_{new} < \chi^2_{old}$,
the new `trial' configuration of the velocity field is accepted.
If $\Delta {\cal L} > 0$, the trial configuration is accepted with a 
probability 
$P_{\Delta {\cal L}} = \exp( - \frac{1}{2}\Delta {\cal L})$, i.e. we take a random number
$\zeta$ uniformly distributed in (0,1), and if
$\zeta < P_{\Delta {\cal L}}$, the new configuration is
accepted. If $\zeta > P_{\Delta {\cal L}}$, the change is
rejected and the configuration of the velocity field is returned
back to its previous state. 
Then we repeat the procedure from
step ({\bf 4}) up to $\beta_1$ times to find the best solution (i.e. the
smallest $\chi^2$ value) for the ${\hat \theta}$ under consideration.
$\beta_1$ is chosen such that each $v_j$ has been changed on average
so often that Var($v_j) \sim \sigma^2_t$. -- The procedure is stopped when
$\chi^2 \la \chi^2_{\nu,\alpha}$.

\smallskip\noindent
{\bf 6}) An element $\theta_i$ of ${\hat \theta}$ is chosen at random,
a random change $\Delta \theta_i$ is given to it, and the
procedure starting from step ({\bf 3}) is repeated
(up to $\beta_2$ times) until $\chi^2_{min}$ will be less than an expected value 
$\chi^2_{\nu,\alpha}$. 
If after $\beta_2$ repetitions the condition $\chi^2_{min} < \chi^2_{\nu,\alpha}$
is not satisfied the procedure stops. $\beta_2$ is chosen such that each $\theta_i$ has been
changed on average no less than 30 times to get a representative statistics.

Note that the convergence of the scheme depends on the size
of the simulation box.
The necessary size is unknown {\it a priori}.
If it is too large, the computing time increases considerably, and if it is too
small, the required $\chi^2$ value may not be reached at all. Therefore 
the box size must be chosen with some care 
and should be adjusted to the experimental data.

\subsection{Validity of the RMC simulations}

In order to test the validity of our RMC-computer code, we performed several
tests in which the `observational' H+D Ly$\alpha$ profile was generated
for a given set of the parameters ${\hat \theta} \equiv $
\{~$N_{\rm HI} = 2.5\times10^{17}\ {\rm cm}^{-2}$, D/H = $1.5\times10^{-4}$,
$T_{kin} = 1250$ K, $\sigma_t/v^{\rm H}_{th} = 2.0$, $L/l = 5.0$~\}, and for one random
realization of the one-dimensional velocity distribution. The chosen parameters
are typical for the warm diffuse interstellar gas (cf.  Linsky {\it
et al.} 1995) except for the D/H ratio which was taken approximately 10 times
larger than the mean ISM value 
in order to make the deuterium absorption being more
pronounced in the simulated spectrum.

The test was conducted with the aim of evaluating whether the code is capable
of finding the correct ${\hat \theta}$ and the correct structure of the
stochastic velocity field. 
At first we considered a very soft constraint on the possible shape of the $v(s)$
distribution. Namely, only the blue wing of the H+D Ly$\alpha$ blend
was used in the minimization procedure (this may correspond to the case when the
red wing of the hydrogen profile is distorted by the Ly$\alpha$ forest lines). 
We will see later (Section 3) that the
uncertainty range of the D/H ratio determination found under this 
soft constraint may be considerably reduced
by accounting for the shape of the whole H+D Ly$\alpha$ absorption. 

\begin{figure}
\vspace{-0.2cm}
\centerline{\vbox{
\psfig{figure=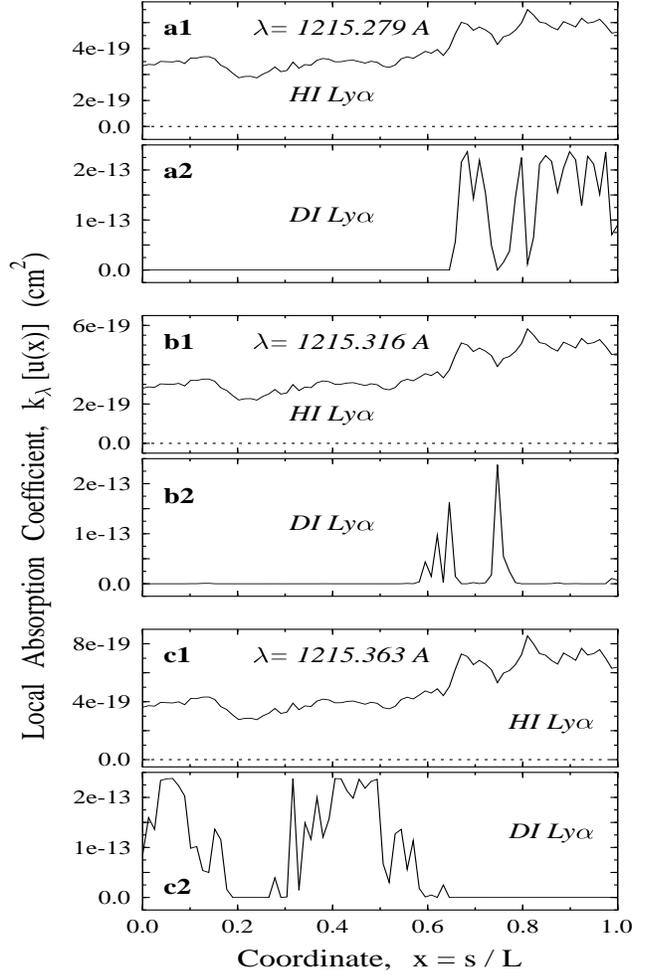,height=15.5cm,width=12.0cm}
}}
\vspace{-2.0cm}
\caption[]{
The contribution from different volume
elements to the HI and DI local absorption coefficients for three fixed wavelengths
$\lambda = 1215.279$ \AA\ ({\bf a1,a2}), 1215.316 \AA\ ({\bf b1,b2}), and 
1215.363 \AA\ ({\bf c1,c2}), which correspond to 
the two minima and the maximum inbetween
in the DI profile shown in Fig.~1a. 
}
\end{figure}

Fig.~1a shows our simulated spectrum of H+D Ly$\alpha$. The example exhibits
a complex intensity structure in DI. This complexity is caused by
the finite correlation length effect only and, hence, it should not be
interpreted as being caused by several (at least three) independent cloudlets
with different radial velocities and different 
physical parameters. The velocity field distribution yielding
this spectrum is shown in Fig.~1c by the dotted curve. 

In order to illustrate the effects the correlated structure of the velocity 
field has on the line forming process,
Fig.~2 shows the local
absorption coefficient $k_{\lambda}(s)$ for HI and DI Ly$\alpha$ for three fixed wavelength
$\lambda \simeq 1215.28, 1215.32$ and 1215.36 \AA\  which correspond to 
the two minima and the maximum between them in the DI profile
shown in Fig.~1a. 

The correlations have the effect that for a given value of $\lambda$ the absorption
coefficient may vary strongly along the line of sight. At the wavelengths chosen
this effect is most pronounced for DI. This is in strong contrast to the 
microturbulent model (having the same $\sigma_t$ and $T_{kin}$) in which 
the whole region along the line of sight would 
evenly contribute to the absorption at a given $\lambda$.
This is clearly seen from 
the hydrogen absorption coefficients 
which at the chosen wavelengths, lie in the range of the damping
wing where both micro- and mesoturbulent models have the same
solutions (see Paper II).

\begin{figure}
\vspace{1.5cm}
\centerline{\vbox{
\psfig{figure=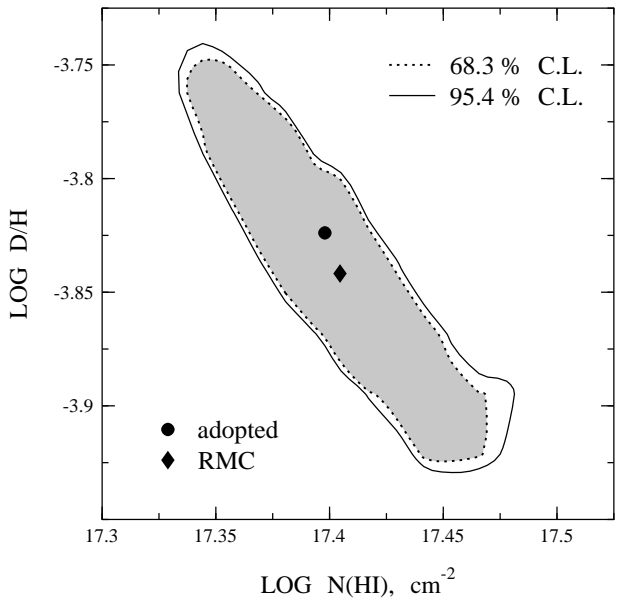,height=12.0cm,width=12.0cm}
}}
\vspace{-5.3cm}
\caption[]{
Confidence region for the plane `log N(HI) -- log D/H'
for the fitted profile from Fig.~1b
when the other parameters
$T_{kin}$, $\sigma_t/v^{\rm H}_{th}$, and $L/l$ are fixed 
but the configuration of the velocity field is free.
The adopted parameters and those found by the RMC procedure
are labeled by the filled circle and diamond, respectively.
}
\end{figure}

Let us turn back to Fig.~1.  The thicker solid
curve in panel {\bf a}  marks the portion of the H+D Ly$\alpha$ profile used in the 
minimization procedure.
To run the RMC-code we sampled the intensity at evenly spaced intervals
$\Delta \lambda$ and added to the $I(\lambda_i)$
the `experimental' errors which correspond to a
signal-to-noise ratio S/N = 30 at the continuum level. 
The corresponding intensities  and 
their uncertainties are shown in Fig.~1b by the open
circles and error bars. 
We consider these data as being `observational' intensities
and apply our RMC-code to estimate  the set of $n = 5$ parameters 
\{ ${\theta}^{\star}_j$ \}
and the corresponding velocity field configuration 
\{ $v^{\star}_j$ \}.
 
The estimated model
parameters are accepted if $\chi^2_{min}$ is found to rank in the
$\chi^2_{\nu,\alpha}$ distribution, i.e. $\chi^2_{min} \simeq 1$. The smooth curve
in Fig.~1b shows one of the possible RMC solutions which gives an adequate fit with
$\chi^2_{min} = 1.188$ (cf.  $\chi^2_{\nu,\alpha} = 1.459$ for
$\nu = 30$ and $\alpha = 0.05$). It corresponds to the
model parameters  
$N^{\star}_{\rm HI} = 2.54\times10^{17}\ {\rm cm}^{-2}$,
(D/H)$^{\star} = 1.44\times10^{-4}$, $T^{\star}_{kin} = 1216$ K,
$(\sigma_t/v^{\rm H}_{th})^{\star} = 2.02$, $(L/l)^{\star} = 5.06$, and the 
$v^{\star}$-configuration shown in Fig.~1c by the solid curve. It is worthwhile
to emphasize once more that the derived $v^{\star}$-configuration is not unique~:
depending on the initial condition (configuration of the velocity field) we may
obtain several patterns for the velocity field distribution that satisfy our
acceptance condition (each of them has its mirror counterpart which is also
acceptable, of course).
But the projected velocity distribution $p(u)$ has approximately the same shape for
all possible RMC solutions. An example of $p(u)$ for 
the adopted and estimated velocity
distributions is shown in Fig.~1d by the dotted and solid histograms, respectively
[here $u$ means the normalized velocity $u(s) = v(s)/\sigma_t$].

The RMC procedure may result in the velocity field configuration becoming
stuck in a local minimum of the ${\cal L}$-function for a very long period 
(`meta-stable' state). Such a minimum may not solve the required problem at all.
This difficulty may be circumvented by making a few starts from different
random initial configurations.
Numerical experiments have shown that for a given
point ${\hat \theta}$\, 
the $\chi^2_{min}$ values may scatter by about 30 per cent around the mean due to
the probabilistic character of the solution for the velocity field. 
This fact should be taken into account
when the error surface is calculated in order to determine the confidence regions for
the estimated parameters 
\{ $\theta^{\star}_j$ \}. To get meaningful confidence limits
for fitted parameters the error surface must be calculated several times  
with different starting points.
Having calculated a set of error
surfaces, the best one may be obtained by 
simply picking  the smallest $\chi^2$ values
for each point in the computational grid.  

\begin{figure}
\vspace{1.5cm}
\centerline{\vbox{
\psfig{figure=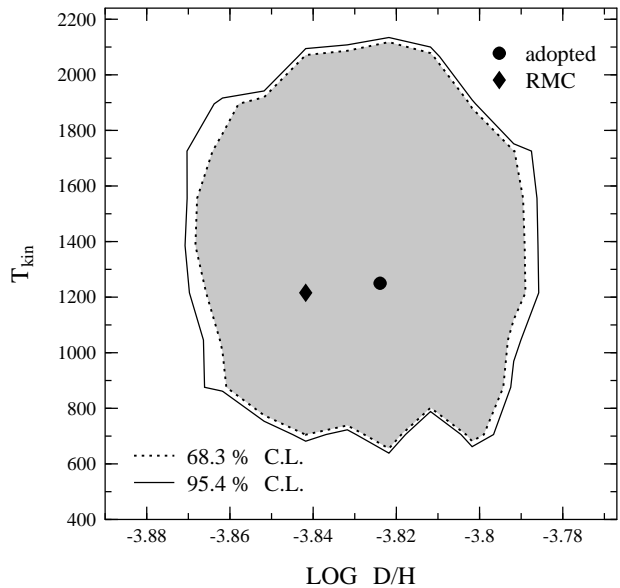,height=12.0cm,width=12.0cm}
}}
\vspace{-5.3cm}
\caption[]{
Confidence region for the plane `log D/H -- $T_{kin}$'
for the fitted profile from Fig.~1b
when the other parameters
$N_{\rm HI}$, $\sigma_t/v^{\rm H}_{th}$, and $L/l$ are fixed 
but the configuration of the velocity field is free.
The adopted parameters and those found by the RMC procedure
are labeled by the filled circle and diamond, respectively.
}
\end{figure}

Two examples of the confidence regions for the `log $N_{\rm HI}$ -- log D/H' and
`log D/H -- $T_{kin}$'  
projections are presented in Fig.~3 and 4, respectively.
The results were obtained from a set of six computed error surfaces for each case.
Shown are two confidence  `ellipses' -- the regions
constrained by the $\chi^2$ values corresponding to 
68.3~\%  and 95.4~\% 
C.L. (1$\sigma$  and 2$\sigma$ regions for a normal distribution
of residues, respectively).  
The RMC solutions are labeled by the filled diamonds whereas the filled circles mark the
adopted quantities used to produce the `observed' spectrum.
A good agreement within statistical errors between
these two points and between the two projected velocity field distributions 
can be considered as a justification of our computational procedure.

It is worth emphasizing that the elongated and declined shape of the confidence regions 
in Fig.~3 
shows that the D/H value is strongly anti-correlated with $N_{\rm HI}$.
Therefore the D/H estimates
may be badly dependent on the assumed hydrogen column density.
 
The correlation between $T_{kin}$ and D/H is less pronounced.
However, Fig.~4 shows that the kinetic temperature 
is not well determined in general.
It seems reasonable to assume that the narrowest subcomponent of a complex absorption
spectrum can show only an upper limit for $T_{kin}$ since the 
correlated structure of the velocity field hampers the exact measurement
of $T_{kin}$ from the apparent $b$-values (see Paper~II, for details).

\section{Application to QSO data}

The example from Subsection 2.2 demonstrates that complexity 
(and/or asymmetry) of an observed absorption spectrum 
is not necessary an indication of a clumpy structure of the absorbing material but
may be solely caused by the correlations in the velocity field.
Therefore one may try to explain the observed spectra starting with a minimum number
of {\it ad hoc} assumptions concerning the structure of the absorber.
In particular, one may consider a cloud model with a homogeneous density
and temperature, assuming that the observed H+D absorption from QSO spectra
arises in the outer regions (halos) of the putative intervening galaxies
(cf. Tytler {\it et al.} 1996).
If this is the case, then $N_{\rm HI}$ ranges from $\sim 10^{17}$ to $\sim 10^{18}$
cm$^{-2}$. This in turn means that the geometrical size of the absorbing material
(parameter $L$ in our model) may be very large, $L \sim 10$ kpc. If there are
many random velocity elements along the line of sight ($L/l \gg 1$), the theoretical
expectation value $\langle I_\lambda \rangle$ should be a good approximation to the
observations (Paper I). If, however, the ratio $L/l$ is not too large, deviations
of $\langle I_\lambda \rangle$ from the observed profile (different kinds of asymmetry)
are to be expected. 
 
Modern observations of the H+D absorption at high redshift revealed two
limiting D/H ratios $ \sim 2\times10^{-4}$
(e.g. Carswell {\it et al.} 1996), and $\sim 2\times10^{-5}$ (e.g. Tytler {\it et al.} 1996). 
The primordial D/H ratio
is shown to be very sensitive to the physical conditions that existed at the
time of the Big Bang (e.g. Walker {\it et al.} 1991). D/H is expected
to decrease with cosmic time due to processing of gas by stars. 
The theoretical uncertainties in the
calculations of the primordial 
hydrogen isotopic ratio are rather small, the D/H value being
predicted for a given baryon-photon ratio, $\eta$, 
with an accuracy of $\simeq 15$ \% (see Sarkar 1996). 
Therefore, accurate measurements of the primordial D/H ratio would provide important
constraints on the cosmological theories.
Since these measurements are of fundamental
importance to cosmology, the validity of the revealed two limiting
D abundances and the reliability of the quoted errors should be
thoroughly investigated.

In the following subsection we will show that accounting of  
the velocity field structure of the large scale gas flows
may have a major impact to the D/H measurements.

\begin{figure*}
\vspace{-1.0cm}
\centerline{\vbox{
\psfig{figure=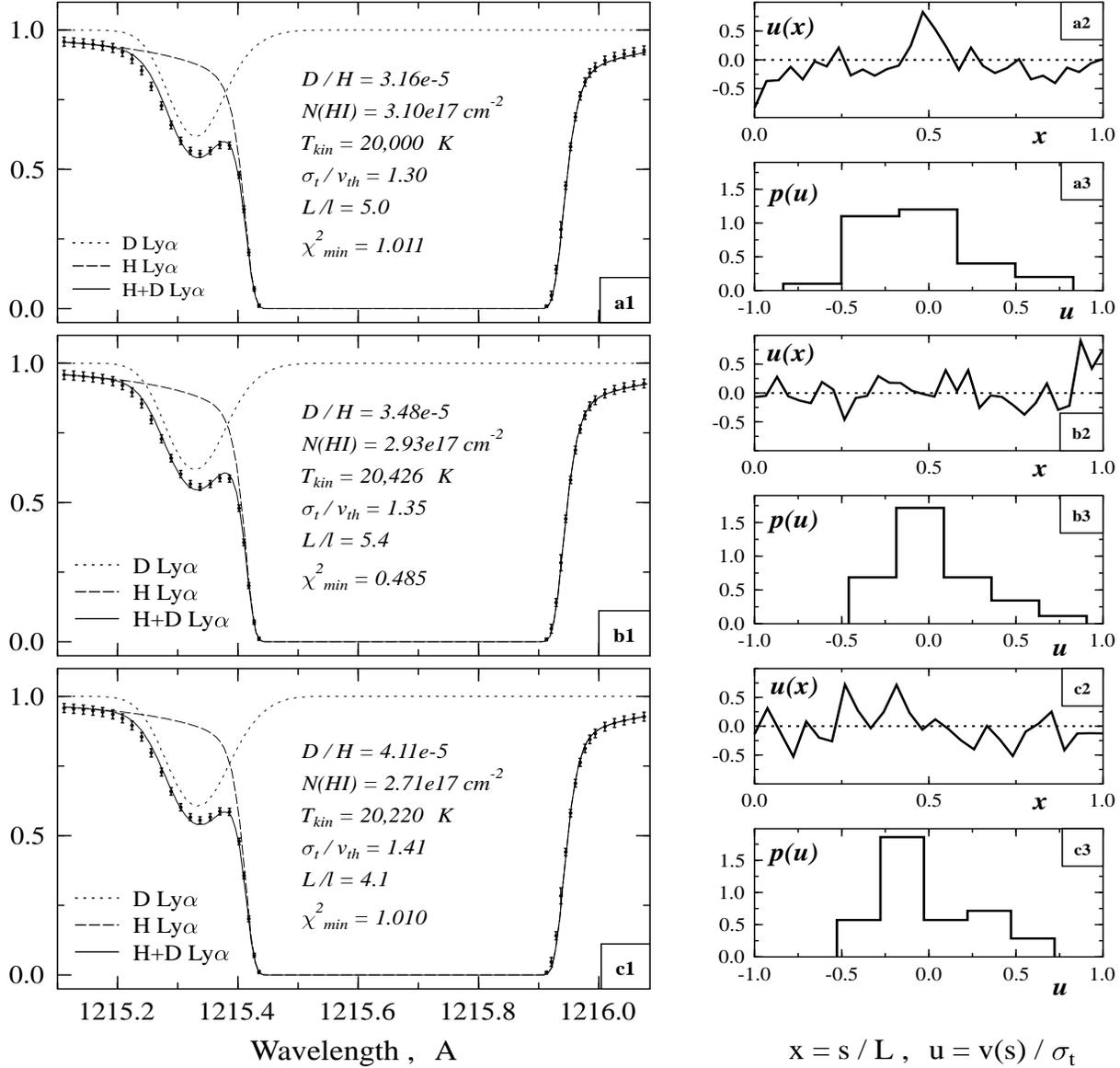,height=17.0cm,width=16.0cm}
}}
\vspace{0.0cm}
\caption[]{
({\bf a1, b1, c1}) --
A template H+D Ly$\alpha$ profile (dots with error bars) representing
the normalized intensities and their uncertainties in accord
with the B\&T data. The solid curves show 
the results of the RMC minimization, whereas the dotted and dashed curves are
the separate profiles of DI and HI, respectively.
Also shown are the best fitting parameters and the $\chi^2$ values obtained for each case.
({\bf a2, b2, c2}) -- The corresponding individual realizations of the 
one-dimensional velocity distributions $u(x)$ in units of $\sigma_t$.
({\bf a3, b3, c3}) -- The histograms are the projected velocity distributions
$p(u)$. 
}
\end{figure*}

\subsection{The $z_a = 2.504$ system in Q 1009+2956} 

The RMC procedure was applied  to 
the D absorption observed by B\&T at $z_a = 2.504$ in front of
the quasar Q 1009+2956.

Using a {\it two}-cloud microturbulent model
and applying a Voigt profile fitting procedure,
B\&T derived the following parameters for this system~: total hydrogen
column density $N_{\rm HI}$ = $2.9^{+0.4}_{-0.3} \times 10^{17}$
cm$^{-2}$ and D/H = $3.0^{+0.6}_{-0.5} \times 10^{-5}$; individual
$T_{kin}$ = $2.1^{+0.1}_{-0.1} \times 10^4$ K \, and
$2.4^{+0.7}_{-0.7} \times 10^4$ K for the blue and red
subcomponents, respectively, the corresponding turbulent
velocities $b_{turb} = 3.2 \pm 0.4$ km s$^{-1}$ and
$2.3 \pm 1.4$ km s$^{-1}$, and a difference in the radial velocity
(derived from metal lines) of 11 km s$^{-1}$.
Including corrections for possible weak HI absorption at the position of D
(estimated as an expected value in Monte Carlo simulations),
B\&T reduced the measured D/H ratio to 
log (D/H) = $-4.60 \pm 0.08 \pm 0.06$ (i.e. D/H $\simeq 2.5\times10^{-5}$),
where the $1\sigma$ random error followed by the systematic error from
fitting the continuum level are given. However, for a given line
of sight the proper correction is never known,  
and therefore we will not reduce 
D/H values in our analysis. This should be borne in mind when discussing the
true D abundance.

We start with the calculation of a template spectrum using the parameters
of B\&T listed in their Table~1.
To simulate real data, 
the experimental uncertainties are added to the computed intensities
which were sampled in equidistant bins as shown in 
Fig.~5({\bf a1},{\bf b1},{\bf c1}) by dots and corresponding error bars. 
For the redshift the mean value of 2.504 was adopted and a
{\it one}-component mesoturbulent model with a
homogeneous density and temperature was used to construct the objective function.

Adequate profile fits ($\chi^2_{min}$ per degree of freedom $\la 1$)  
for three different sets of parameters are
shown in panels ({\bf a1},{\bf b1},{\bf c1}) of Fig.~5 by solid curves, 
whereas the individual HI and DI Ly$\alpha$ profiles
are shown by dashed and dotted curves, respectively.
The estimated parameters and corresponding $\chi^2_{min}$ values
are also listed in these panels for each solution.
The adjusted $u(x)$-configurations (panels {\bf a2},{\bf b2},{\bf c2}) 
and their projected distributions $p(u)$ (panels {\bf a3},{\bf b3},{\bf c3})
show pronounced deviations from a single Gaussian. 
Such distributions may, in principle, have very long positive `tails'
yielding unobservable D absorption which is hidden by the core of the hydrogen line.
However, including in the objective function both
the blue and the red wing of the HI line, allows to constrain 
significantly the set of the $u(x)$-configurations. As a result
the uncertainty for the D/H measurements 
is considerably reduced.

In accord with our calculations, the restored profile of the DI line in
the $z_a = 2.504$ system is almost symmetrical. A slight asymmetry of
the red DI Ly$\alpha$ wing is caused by the weak distortion of the $p(u)$
distribution toward positive $u$-values. In this case the width of the DI line
may be formally characterized by the $b$-parameter~:
$b_{\rm DI} = v^{\rm D}_{th} [ 1 + 2(\sigma_t/v^{\rm D}_{th})^2 ]^{1/2}$ .
The inspection of Fig.~5 ({\bf a1},{\bf b1},{\bf c1}) reveals rather small $b_{\rm DI}$
value ($\simeq 15$ km s$^{-1}$) as compared with $\sigma_t \simeq 25$ km s$^{-1}$
obtained by the RMC procedure. This is not a surprise, however, because random
realization of the correlated velocity field may have  $p(u)$ which deviates
substantially from the ensemble average distribution (assumed to be Gaussian in this case)
leading to a `contraction' of this  distribution (see Paper II).

One may further be interested in the question whether such weak asymmetry of $p(u)$ 
is essential for the analysis of the H+D Ly$\alpha$ profile, 
or whether a simpler
one component microturbulent model could be used to study
the $z_a = 2.504$ absorber.  The answer is that it is essential indeed, and that the simplified
model fails to reproduce the profile in question. The smallest $\chi^2_{min}$ value
obtained for this model is only 2.4 which rejects the 
one-component microturbulent solution.
This example clearly demonstrates once more that the analysis of absorption spectra
may be very sensitive to the model assumptions concerning the velocity field structure.

The essential difference between the results of B\&T and ours lies in the estimation
of the hydrodynamical velocities in the $z_a = 2.504$ absorbing region.
The two-component microturbulent model yielded
$b^{\rm D}_{blue} = b^{\rm D}_{red} = 14$ km s$^{-1}$ and $b_{turb} \simeq 3$ km s$^{-1}$ 
which is about 10 times smaller than $\sqrt{2}\sigma_t$ estimated by the
RMC procedure. For the additional parameter $L/l$ of the mesoturbulent model,
we found a value of $\simeq 5$, indicating 
that the space averaging along the line of sight corresponds to an
incomplete statistical sample and, hence, significant deviations of
the observed intensities from their expectation values may occur.

As for the other parameters ($N_{\rm HI}$, D/H, and $T_{kin}$), the RMC method yields
in this particular case values which are similar to those of B\&T. The total
hydrogen column densities and the D/H ratios obtained are shown in Fig.~6 by
the filled diamond for the best RMC solution  (panel {\bf b1} in Fig.~5) and by the filled
circle for the B\&T model (the error bars shown correspond to $1\sigma$ in accord with
the B\&T data from Table~1). The filled triangle and square mark the ($N_{\rm HI}$, D/H)  
pairs found separately for the blue and red components. The D abundance for the
red component was assumed by B\&T as 
(D/H)$_{red}$ = (D/H)$_{blue}$ = (D/H)$_{total}$  and, thus,
there was no real measurements of the D/H value for the red component in their model.
This is marked by arrows instead of error bars at the filled square in Fig.~6.
The projection of the 5-dimensional error surface onto the 
`log $N_{\rm HI}$ -- log D/H' plane was calculated by the RMC procedure using 
the $\Delta (\chi^2)$ method. Drawn are 68 \% and 95 \% confidence regions
computed under the condition that the  parameters $T_{kin}$, 
$\sigma_t / v_{th}$, and $L/l$ are {\it fixed}, but the velocity field configuration
is {\it free}. As seen, the B\&T value for the total $N_{\rm HI}$ is located closely to the
RMC confidence regions, but the D/H ratio is systematically lower in the
microturbulent model. 

Fig.~7 shows 68~\% and 95~\% confidence regions for the two parameters
$T_{kin}$ and D/H computed
under the assumption that $N_{\rm HI}$, $\sigma_t/v_{th}$, and $L/l$ are fixed, but the
$u(x)$-configuration is free. The kinetic temperature is found in the range from
$\sim 15000$ K to $\sim 22000$ K, whereas D/H lies between $\sim 3.3\times10^{-5}$
and $\sim 4.5\times10^{-5}$. In this figure, the filled symbols label the RMC and B\&T
best fitting parameters in the same sense as in Fig.~6. Note the marginal
position of the RMC solution (the filled diamond) which shows that there are points
in the `D/H -- $T_{kin}$' plane with $\chi^2$ values less than 
$\chi^2_{min} = 0.485$ adopted for the best RMC solution at the first stage of calculations. 
Fig.~7 shows that also in the `D/H -- $T_{kin}$' plane 
the D/H uncertainty region is shifted 
toward higher D/H values as compared with the B\&T data. 

\begin{figure}
\vspace{1.5cm}
\centerline{\vbox{
\psfig{figure=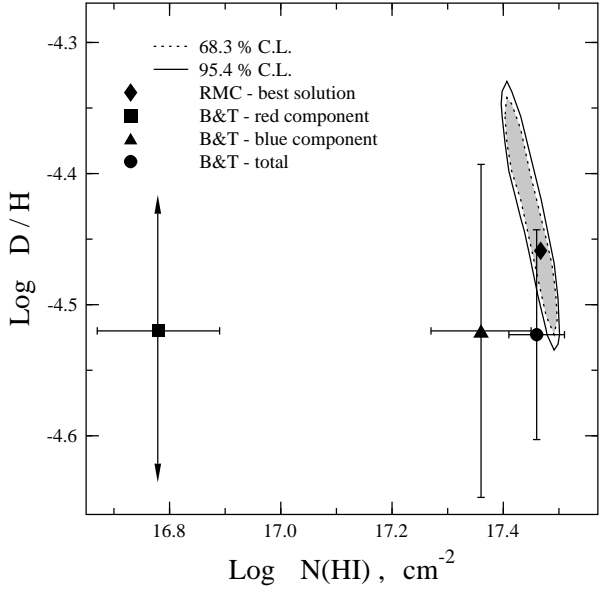,height=12.0cm,width=12.0cm}
}}
\vspace{-5.3cm}
\caption[]{
Confidence region for the plane `log N(HI) -- log D/H'
for the fitted profile from Fig.~5(b1)
when the other parameters
$T_{kin}$, $\sigma_t/v^{\rm H}_{th}$, and $L/l$ are fixed 
but the configuration of the velocity field is free.
The RMC solution is labeled by the filled diamond. The
filled square and triangle represent the B\&T parameters
for the red and blue components of their two-component
microturbulent model, and the filled circle marks  the
combined solution. The error bars correspond to 1$\sigma$
in accord with B\&T, but the arrows reflect the fact that
the D abundance was estimated through the blue component mainly
and (D/H)$_{red}$ = (D/H)$_{blue}$ was assumed by B\&T.
}
\end{figure}

\begin{figure}
\vspace{1.5cm}
\centerline{\vbox{
\psfig{figure=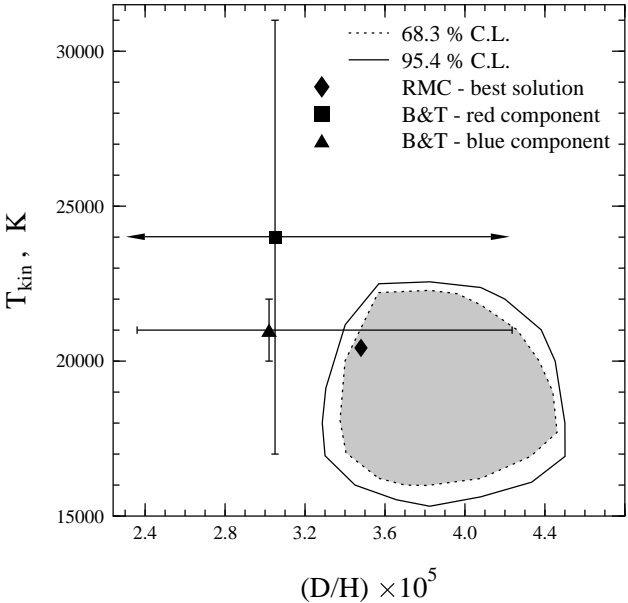,height=12.0cm,width=12.0cm}
}}
\vspace{-5.3cm}
\caption[]{
Confidence region for the plane `D/H -- $T_{kin}$'
for the fitted profile from Fig.~5(b1)
when the other parameters
$N_{\rm HI}$, $\sigma_t/v^{\rm H}_{th}$, and $L/l$ are fixed 
but the configuration of the velocity field is free.
The symbols used have the same meaning as in Fig.~6.
}
\end{figure}

Above it was noted that the D/H ratio does sensitively 
depend on the $N_{\rm HI}$ value chosen.
The measured D/H value 
and its confidence interval are both also affected by the choice
of the velocity field configuration. 
The accuracy of the RMC solution may be increased if some
additional information about the velocity field structure 
is available for the analysis. 
For instance, the profile shapes of higher order Lyman lines 
may yield additional useful constraints to the allowable $p(u)$ distribution.
For this case the size of the D/H uncertainty interval may be narrowed down.
This is demonstrated in Fig.~8 where 
the results of two calculations (formally called `high' and `low' D/H) are shown.
The 68~\% and 95~\% confidence regions were computed for the {\it fixed}
$T_{kin}$, $\sigma_t/v_{th}$, and $L/l$ listed in panels {\bf c1} 
and {\bf a1} (Fig.~5), respectively,
and for the {\it fixed} configurations of $u(x)$ shown in panels c2 and a2 (Fig.~5).
Provided the constraints mentioned above are available and, thus
a distinction between different $p(u)$ distributions compatible with the measured
H+D profile (Fig.~5) is possible,
the accuracy of the D/H measurements improves essentially becoming comparable
with the accuracy of the SBBN calculations (Sarkar 1996).

We now discuss the observed D abundance and its inferred
primordial value which can be compared with the SBBN model predictions.
Within the framework of the SBBN, the primordial abundances of the light
elements (D, $^3$He, $^4$He, $^7$Li) depend only on the baryon to photon
ratio $\eta = n_b / n_{\gamma}$, where 
$n_b$ is the current cosmological number density
of baryons and $n_{\gamma}$ is the number density
of the cosmic background photons.
The predicted relations between the abundances of D, $^4$He, and $^7$Li 
and $\eta$  are illustrated in Fig.~9. The thick solid curves plotted in this 
figure are computed by using the analytical formulae from Sarkar (1996), while
dashed curves depict their uncertainty intervals. The primordial mass fraction of
$^4$He, $Y_p = 0.243 \pm 0.003$, was derived by Izotov {\it et al.} (1997),
and Bonifacio \& Molaro (1997) give the following value for the lithium
abundance in halo population II stars~: log ($^7$Li/H)$_p^{\rm II}$ =
$ -9.762 \pm 0.012 \pm 0.05$ (here the subscript p denotes primordial).
The uncertainty regions for the measured values of Y$_p({}^4$He) and  
($^7$Li/H)$_p^{\rm II}$ and the implied SBBN limits on $\eta$ are shown by
the solid-line rectangles in Fig.~9(b,c). The
upper panel {\bf a} in this figure gives our results~: 
the solid-line rectangle is defined by
the 95~\% confidence region (shown in Fig.~6) for an arbitrary configuration of the
velocity field, whereas two dotted-line rectangles 
correspond to 68~\% confidence regions computed for the two fixed velocity fields
leading to the `high D/H' and `low D/H' solutions shown in  Fig.~8.
The shaded region in Fig.~9 represents the range of $\eta$ 
compatible with the measured abundances of all three nuclei.
Fig.~9 shows that concordance of D/H with Y$_p({}^4$He) and 
($^7$Li/H)$_p^{\rm II}$ measurements is achieved if 
$3.9\times10^{-10} \la \eta \la 5.3\times10^{-10}$. 
This constraint on $\eta$ is consistent with results of other groups; for
example, Hata {\it et al.} (1995) quote $\eta = 4.4^{+0.8}_{-0.6}\times10^{-10}$ $(1\sigma)$
[see also Sarkar (1996) for a discussion of the intercomparisons of the allowed
$\eta$-intervals]. The center of the shaded $\eta$-interval in Fig.~9
corresponds to $\Omega_b h^2 = 0.017$, where $\Omega_b$ is the fraction of the
critical density contributed by baryons and $h$ the Hubble constant in units 
100 km s$^{-1}$ Mpc$^{-1}$. The uncertainties in this estimation can alter $\Omega_b$
by up to $\simeq 20$~\%.

\begin{figure}
\vspace{1.5cm}
\centerline{\vbox{
\psfig{figure=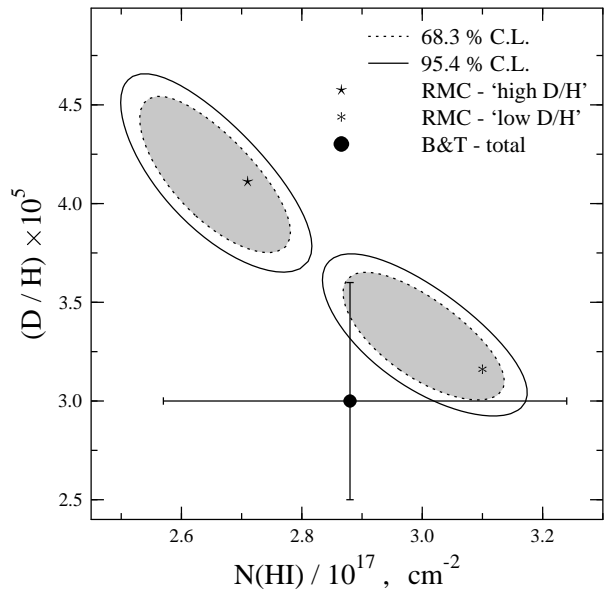,height=12.0cm,width=12.0cm}
}}
\vspace{-5.3cm}
\caption[]{
Confidence region for the plane `N(HI) -- D/H'
for the fitted profiles from Fig.~5 {\bf a1} and {\bf c1}
when the other parameters
$T_{kin}$, $\sigma_t/v^{\rm H}_{th}$, and $L/l$ and
the corresponding configurations of the velocity fields
(shown in Fig.~5 {\bf a2}, {\bf c2}) are fixed.
The RMC `high D/H' and `low D/H' solutions are labeled by star and
asterisk, respectively. The filled circle and 1$\sigma$ error bars represent
the B\&T result.
}
\end{figure}

\section{Conclusions}

The origin of QSO absorption-line systems with HI column densities
$\sim 10^{17} - 10^{18}$ cm$^{-2}$ (which are particularly useful for
deuterium observations) is usually associated with extended gas halos 
of foreground galaxies. 
According to recent high resolution observations, 
the line profiles in these systems
are often wider than expected from purely thermal broadening, and,
therefore, their widths reflect
both thermal and turbulent broadening caused by any kind of bulk motions
in the halos. 
In turbulent media the determination of HI and DI
column densities depends on the model (micro- or mesoturbulent) 
is chosen for the profile analysis. 
The  microturbulent model yields reasonable results if
$\sigma_t \ll v_{th}$ , or $L/l \gg 1$, otherwise the mesoturbulent one 
is more adequate.  

\begin{figure}
\vspace{0.6cm}
\centerline{\vbox{
\psfig{figure=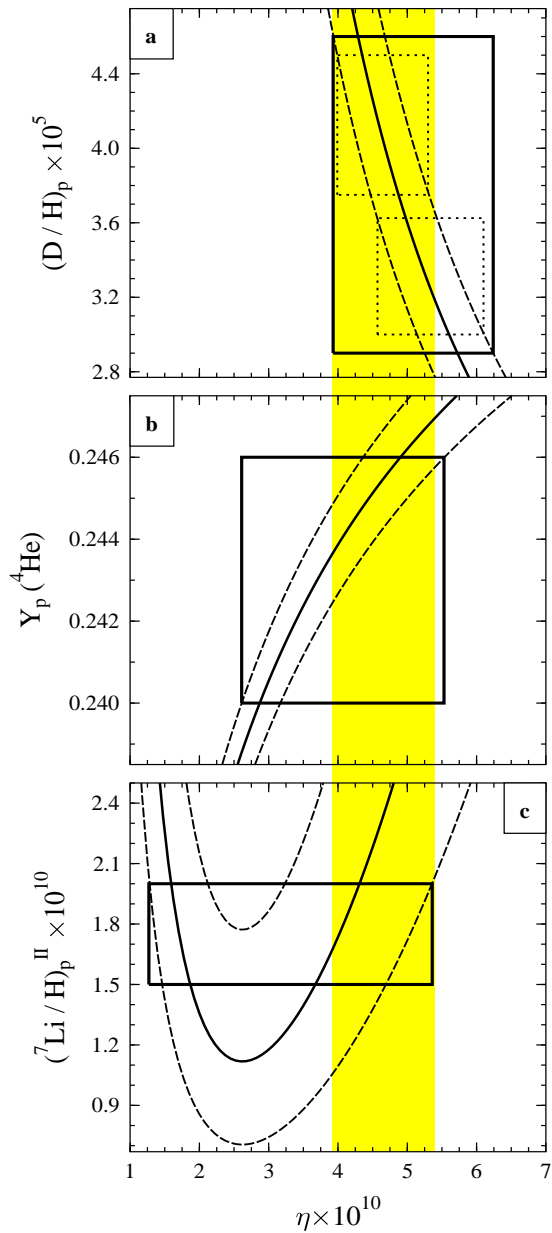,height=15.0cm,width=12.0cm}
}}
\vspace{1.4cm}
\caption[]{
Comparison of predicted primordial abundances 
(denoted by the subscript p) with observational bounds. 
The theoretical SBBN \ D, $^4$He, and $^7$Li yields (solid curves) and their uncertainties
(dashed curves) as function of $\eta$ (the baryon to photon number ratio) are computed by
using the parameterization of Sarkar (1996). The abundances of D and $^7$Li are number ratios,
whereas Y$_p$ is the mass fraction of $^4$He. 
The hight of the solid-line rectangles in panels
{\bf b} and {\bf c} give the bounds from recent measurements of extra-galactic Y$_p$ 
(Izotov {\it et al.} 1997) and $^7$Li in population II (halo) stars 
(Bonifacio \& Molaro 1997). The horizontal widths give the range of $\eta$ compatible
with the measurements. In panel {\bf a} the
solid-line rectangle corresponds to the 2$\sigma$ uncertainty region for D/H towards Q1009+2956,
as calculated by the RMC procedure for an arbitrary velocity field configuration (see Fig.~6),
whereas the two dotted-line rectangles are defined by the 68~\% confidence regions shown in 
Fig.~8 (the velocity field structure is fixed). 
The shaded region is the window for $\eta$ common to all results. 
}
\end{figure}

Direct observations of the Ly$\alpha$ emission 
of distant galaxies ($z > 2$) show that their giant halos 
($R \sim 100$ kpc) have complex morphologies and kinematics
(R\"ottgering {\it et al.} 1996).
The observations reveal as well extended HI absorption systems
(with projected sizes up to $\sim 50$ kpc)  
seen against the Ly$\alpha$ emission (van Ojik {\it et al.} 1997). 

Using the measured Doppler parameters for the absorbing gas
with $N_{\rm HI} \simeq 10^{18.1} - 10^{18.5}$ cm$^{-2}$
(see Table~3 in van Ojik {\it et al.}) and the kinetic temperature $T_{kin} = 10^4$ K
 adopted by van Ojik {\it et al.}, we can estimate the range
for the velocity dispersion of bulk motions~: 
24 km s$^{-1} \la \sigma_t \la 55$ km s$^{-1}$.
It follows that $\sigma_t/v_{th} > 1$ within the HI absorption line gas embedded
in the galactic halos. Both findings -- high absolute value of $\sigma_t$ and
high $\sigma_t/v_{th}$ ratio, -- make the mesoturbulent approach to appear
to be more realistic one.
Besides, our present RMC calculations yielded for the $z_a = 2.504$ absorption system 
$\sigma_t \simeq 25$ km s$^{-1}$
which lies just within the observed range, whereas $\sigma_t \simeq 2$ km s$^{-1}$
found by B\&T is evidently too low.

The main conclusion of our work is that
the mesoturbulent approach together with 
the RMC computational scheme 
allows to solve the inverse problem for the H+D Ly$\alpha$ absorption
and to restore the physical parameters of the absorbing gas
as well as the projection of the velocity field distribution.

The proposed computational procedure enables us to draw
confidence regions for different pairs of the adopted parameters.
From the analysis of the synthetic H+D Ly$\alpha$ spectrum
it was found that, in general, the measured D/H and $N_{\rm HI}$ values
are anti-correlated. 

The study of the template H+D Ly$\alpha$ profile (which reproduces
the original spectrum of B\&T) yields D/H = $(3.75 \pm 0.85)\times10^{-5}$ $(2\sigma)$.
We therefore adopt the value of
$4.6\times10^{-5}$ as a conservative upper limit on the primordial abundance of
D relative to hydrogen.

We conclude that the discordance of D/H with the $^4$He and $^7$Li primordial abundances
noted by B\&T is a consequence of the use of the microturbulent model. 
Within the framework of the generalized model one finds  good
agreement between the measurements mentioned above and the SBBN predictions.

\subsection*{Acknowledgments}

This work was supported by the Deutsche Forschungsgemeinschaft, 
and by the RFBR grant No. 96-02-16905-a.
The authors thank Dr. I. Agafonova for valuable suggestions
on the RMC technique and for her kind help in the development
of the RMC computer code. 
SAL gratefully acknowledges the hospitality of the
Institut f\"ur Theoretische Physik der Universit\"at Frankfurt am Main.

\end{document}